\documentclass[fleqn,twoside]{article}
\usepackage{lscape}
\usepackage{graphics}
\usepackage{epsfig}
\usepackage{graphicx}
\usepackage{rotate}
\usepackage{color}
\usepackage{espcrc2}

\usepackage{graphicx}
\usepackage[figuresright]{rotating}

\newcommand{\AmS}{{\protect\the\textfont2
  A\kern-.1667em\lower.5ex\hbox{M}\kern-.125emS}}

\def\ApJ{{\it Astrophys. J.} }
\def\ApJL{{\it Astrophys. J. Letters} }

\def\ApP{{\it Astropart. Phys.} }
\def\AA{{\it Astron. \& Astroph.} }

\def\MNRAS{{\it Month. Not. Roy. Astr. Soc.} }
\def\Nature{{\it Nature} }
\def\NewAR{{\it New Astron. Rev.} }

\def\PRL{{\it Phys. Rev. Letters} }

\def\simle{\lower 2pt \hbox {$\buildrel < \over {\scriptstyle \sim }$}}
\def\simge{\lower 2pt \hbox {$\buildrel > \over {\scriptstyle \sim }$}}

\title{Photon and neutrino emission from active galactic nuclei}

\author{Peter L. Biermann\address[MPIfR]{MPI for Radioastronomy, Bonn, Germany}%
        \thanks{PLB also at: Dept. of Phys., Karlsruher Institut f{\"u}r Technologie KIT, Germany; Dept. of Phys. \& Astr., Univ. of Alabama, Tuscaloosa, AL, USA; Dept. of Phys., Univ. of Alabama at Huntsville, AL, USA; Dept. of Phys. \& Astron., Univ. of Bonn, Germany },
        Julia K. Becker \address[Bochum]{Inst. for Phys., Univ. Bochum, Bochum, Germany},
	Lauren\c{t}iu I. Caramete \addressmark[MPIfR]\thanks{LIC main address: Institute for Space Sciences, Bucharest, Romania},
	Federico Fraschetti \addressmark[Bochum]\thanks{FF main address: Lunar and Planetary Lab. \& Dept. Physics-Theor. Astrophys., Univ. of Arizona, Tucson, AZ, USA;  also at LUTh, Observatoire de Paris, CNRS \& Univ. Paris VII, Meudon, France},
		Tanja Kneiske\address{Inst. f{\"u}r Exp.Physik, Univ. Hamburg, Hamburg, Germany},
		Athina Meli\address{Erlangen Center for Astroparticle Physics, University
Erlangen-Nuremberg, Germany},
		Todor Stanev\address{Bartol Research Inst., Univ. of Delaware, Newark, DE, USA}}
       
\begin{document}

\begin{abstract}
Supermassive black holes in the centers of galaxies are very common.  They are known to rotate, accrete, spin down and eject highly relativistic jets; those jets pointed at us  all seem to show a spectrum with two strong bumps, one in the TeV photon range, and one in X-rays - ordered by the emission frequency of the first bump this constitutes the blazar sequence.    Here we wish to explain this sequence as the combined interaction of electrons and protons with the magnetic field and radiation field at the first strong shockwave pattern in the relativistic jet.  With two key assumptions on particle scattering, this concept predicts that the two basic maximum peak frequencies scale with the mass of the central black hole as $M_{BH}^{-1/2}$, have a ratio of $(m_p/m_e)^{3}$, and the luminosities with the mass itself $M_{BH}$. Due to strong losses of the leptons, the peak luminosities are generally the same, but with large variations around equality.  This model predicts large fluxes in ultra high energy cosmic rays, and also large neutrino luminosities.
\vspace{1pc}
\end{abstract}

\maketitle

\section{Introduction}

Every massive galaxy harbors a central supermassive black hole; many of these black holes rotate, accrete, spin down, and eject powerful highly relativistic jets.  These jets are visible through their emission from radio wavelengths to TeV photons, and have a  very knotty structure, interpreted usually as shocks in the flow.  Whenever these jets are pointing in our direction, the emission is extremely enhanced by Lorentz-boosting; we note that this is true for half of all sources in any flux density limited complete sample of sources observed at 5 GHz.  The typical spectrum in many cases includes a bump in TeV photons and another bump in X-rays; the two bumps can appear at a range of frequencies, but appear to range in frequency in parallel.  This sequence is called the ``blazar sequence" \cite{FMCCG98}, and this is what we wish to explain.

We first assume, that we are in the spin-down phase of a black hole of mass $M_{BH}$ with very high spin, that phase, when the rotation of the black hole \cite{BZ77} powers the activity.  This is consistent with data for the radio galaxies M87 and Cen A \cite{WA03}.  This implies that the magnetic fields are accreted during an earlier phase of maximal accretion, and decay extremely slowly.  This gives a jet power $L_j$ of  

\begin{equation}
L_j \; = \; 10^{43.5} \, f_j \, \frac{M_{BH}}{10^{8} \, M_{\odot}} \, {\rm erg/s} 
\end{equation}

with an arbitrary scaling parameter $f_j$ of order unity.  This also gives a scaling of the magnetic field inside the jet  \cite{Love76}, as $B \, \sim \, M_{BH}^{-1/2}$ in the observer frame.  We denote the comoving frame with dashes, as in $B' \, = \, B /\Gamma_j$, where $\Gamma_j$ is the Lorentz factor of the jet flow.

Furthermore we assume, following earlier work \cite{MFF01}, that the first strong shock system is at a fixed relative distance from the black hole about $10^{3.5} $ gravitational radii $r_g \, = G_N M_{BH}/c^{2}$, where $G_N$ is Newton's constant of gravitation. We then work out an acceleration condition, the characteristic frequencies of emission for electrons and protons, and the corresponding luminosities.  A key argument is that the electrons are in the loss-limit \cite{Kardashev62}, and so their synchrotron luminosity is greatly reduced compared to original expectations \cite{BS87}.

\section{Hadrons versus Leptons}

We picture a relativistic flow with a strong oblique stationary
shock \cite{MBQ08}.  A diametrical length scale $z \, \theta$ is tilted and so increased by the Lorentz factor $\Gamma_j$ for a transverse flow component which is only weakly relativistic; $z$ is the distance along the jet, and  $\theta$ is the jet opening angle.  Transforming the corresponding timescale into the jet flow frame gives another factor of $\Gamma_j$.  We write this time scale as $(z \, \Gamma^{2}_j \, \theta \, \epsilon)/(\beta_j c)$, where $\epsilon$ is a parameter of order unity, and $\beta_j c$ the jet flow velocity.  We use the flow time in the moving jet frame as a measure of acceleration \cite{BK79}, and then put it equal to the synchrotron loss time of protons.  Thus, in the comoving frame (dashes) one has

\begin{equation}
\frac{z \, \Gamma^{2}_j \, \theta \, \epsilon}{\beta_j c} \; = \; \frac{6 \pi m_{p}^{3} c}{\sigma_T m_e^{2} {\gamma'}_{p} {B'}^{2}} \quad .
\end{equation}

This is the first key assumption.  The proton maximum Lorentz factor is given by (in the limit of $\Gamma_j >> 1$):

\begin{equation}
\gamma_{p, max} \; = \; 10^{+11.6} \, f_j^{-1} \, \theta_{-1} \, z_{fs,3.5} \, \Gamma_j \, \epsilon^{-1}
\end{equation}

\noindent in the observer frame; $z = 10^{3.5} \, z_{3.5}$, and the opening angle of the jet $\theta$ is in units of 0.1 rad.  The value $\gamma_{p, max}$ is independent of the black hole mass.  The black hole mass is given in units of $10^{8} \, M_{\odot}$ as $M_{BH, 8}$.  This limit is given by losses; a spatial constraint in the spin-down limit can be expressed as $\gamma_{Z, max} \; \simeq \; 10^{9.5} \, Z \, f_{j}^{1/2} \, M_{BH, 8}^{1/2}$, suggesting that heavy nuclei of charge $Z$ or flaring $ 300 \, \simge \, f_j \, > 1$ may be required.  A condition on the factor $f_j$ follows, for the losses to constrain more than the space.  The proton synchrotron emission frequency  $\nu_{syn, p, max}$ is in the observer frame 

\begin{eqnarray}
\nu_{syn, p, max} && \; = \; 10^{+27.6} \, {\rm Hz} \,  M_{BH, 8}^{-1/2} \,\\  && f_j^{-3/2} \, \theta_{-1} \, z_{fs, 3.5} \, \epsilon^{-2}  \quad .\nonumber
\end{eqnarray}

As noted above, we use the assumption that the flow is highly oblique: in such a case  the flow component parallel to the shock normal direction may be sub-relativistic. We assume, that for the relevant cone in phase space the limiting acceleration length scale for protons initiates a Kolmogorov cascade \cite{BS87}.  The acceleration scale for protons is an injection scale also for a turbulent cascade, and at lower energy the scale is smaller by $(({\gamma_e m_e})/({\gamma_p m_p}))^{1/3}$. This is the second key assumption.  So the corresponding electron emission frequency is

\begin{eqnarray}
\nu_{syn, e, max} && \; = \; 10^{+17.7} \, {\rm Hz} \, M_{BH, 8}^{-1/2} \,\\ && f_j^{-3/2} \, \theta_{-1} \, z_{fs, 3.5} \,\epsilon^{-2} \nonumber
\end{eqnarray}

\noindent and $\nu_{syn, p, max}/\nu_{syn, e, max} \, = \, (m_p/m_e)^3$.  We emphasize that here and above the Lorentz factor of the jet flow does not enter.

The synchrotron luminosities $L_p$ of relativistic protons and $L_e$ of relativistic electrons are given by integrating the losses: for the electrons the loss time is very much shorter than the flow time scale, implying that their luminosity is very much reduced as compared to the estimate in \cite{BS87}, where this was then estimated at $L_p/L_e \, \simeq \, m_e/m_p$.  Here we obtain first

\begin{eqnarray}
&& {L_p} \; \sim \,  \int \{ n_{p, 0} \gamma_{p}^{-2} \} \, \{ \gamma_{p}\, m_{p} c^{2} \} \, \\ && \cdot \frac{\sigma_{T} m_{e}^{2} \gamma_{p} B^{2}}{6 \pi m_{p}^{3} c}  \, d \gamma_p \, z^{2} \, \Delta z \, \sim \, M_{BH} \nonumber
\end{eqnarray}

\noindent and then for the ratio

\begin{equation}
\frac{L_p}{L_e} \; \sim \; \frac{n_{p, 0} m_p }{n_{e, 0} m_e } \, \frac{\gamma_{p, max}}{\ln(\gamma_{e,max} / \gamma_{e,min})} \, {\left(\frac{m_{e}}{m_{p}}\right)}^{+3} \quad .
\end{equation}

This ratio works out to be of order unity with large possible excursions.  On occasion the proton peak can be very much larger than the electron peak, possibly explaining orphan flares, e.g., \cite{Acciari09}.   We assume the simple limit of an $E^{-2}$ particle spectrum, which curvature of the shock surface, lack of shock strength, and substructure can modify.  Both protons and electrons interact copiously, distorting the simple spectrum.  Interactions between primary protons and electrons (plus positrons) and the radiation fields produced by their emission can give quadratic and cubic dependencies, just as in multiple Compton orders of straight leptonic emission, \cite{KPT69}.

Therefore the primary luminosities scale as $M_{BH}^{+1}$, the frequencies as $M_{BH}^{-1/2}$, and the frequency ratio is $(m_p/m_e)^3$.  A prediction is that the spectra of the low frequency bump is always primarily steeper $S_{\nu} \, \sim \, {\nu}^{-1}$ than that of the high frequency bump $S_{\nu} \, \sim \, {\nu}^{-1/2}$ in the simple limit of plane-parallel shock acceleration.  This regularity and simplicity is an advantage compared with other model proposals.  The predicted TeV local spectrum is consistent with the data.

These further interactions in turn will produce enormous neutrino fluxes: however, recently AUGER observations \cite{Augerheavy} suggested the presence of the heavy-mass nuclei in the primary flux of UHECRs. The flux of secondary neutrinos produced in the interaction of nuclei with the photon field at the source will modify significantly the predicted neutrino flux.

\section{Conclusions}

Ultra high energy protons, nuclei, as well as electrons emit synchrotron radiation:  We propose that this explains the regularity of the blazar sequence \cite{FMCCG98}.  Numerous further interactions obviously happen for both leptons and protons and the high energy bump may contain components from the emission of heavier nuclei, and the leptons might be all secondary.  A fortiori, as argued for some time, e.g., \cite{GS63,BS87,RB93,MBQ08,Krishna10}, radio galaxies are key sources of ultra high energy cosmic particles.  These particles interact and produce copiously high energy neutrinos.

\vskip0.5cm

PLB wishes to thank F. Aharonian, K. Mannheim, and J.P. Rachen for discussions.

\end{document}